\documentclass[showpacs,preprintnumbers,pre,nofootinbib,onecolumn,12pt]{revtex4}
\usepackage{amsmath}
\usepackage{graphicx}
\usepackage{dcolumn}
\usepackage{bm}

\input{tcilatex}

\begin{document}

\preprint{}
\title{Spatiotemporal Fluctuation Induced Transition in a Tumor Model with
Immune Surveillance}
\author{Wei-Rong Zhong}
\altaffiliation[ ]{Corresponding Author}
\email{wr-zhong@126.com}
\author{Yuan-Zhi Shao}
\author{Zhen-Hui He}
\affiliation{State Key Laboratory of Optoelectronic Materials and Technologies,\\
Department of Physics, Sun Yat-sen University, 510275 Guangzhou, People's
Republic of China}

\begin{abstract}
We report on a simple model of spatial extend anti-tumor system with a
fluctuation in growth rate, which can undergo a nonequilibrium phase
transition. Three states as excited, sub-excited and non-excited states of a
tumor are defined to describe its growth. The multiplicative noise is found
to be double-face: The positive effect on a non-excited tumor and the
negative effect on an excited tumor.
\end{abstract}

\maketitle

In the past decades, many studies have focused on biodynamics [1-4],
specially noise biodynamics [5-9]. More than ever, cancer research is now an
interdisciplinary effort which requires a basic knowledge of commonly used
terms, facts, issues, and concepts. Phase transition of tumor growth induced
by noises is one of the most novel foundations in recent years [10, 11].
However, in all these studies the systems are zero-dimension and
insufficient to describe the real progress in the field of tumor growth,
furthermore at present the space has become a fundamental variable to study
[1, 12, 13].

Chemotherapy and Immunotherapy remain far from good understanding, although
they as a potential practical partnership have attracted numerous attentions
of scientists for at least one decade [14, 15]. Due to the different
responses of tumor cells to chemotherapy and immunotherapy, more recently
Lake and Robinson suggested that there is an interesting and significative
case for combining chemotherapy and immunotherapy in tumor treatment [14].

In this paper, chemotherapy and immunotherapy are joined by a spatial extend
anti-tumor model with three elements, which are (1) a spatiotemporal
fluctuation of growth rate induced by chemotherapy, (2) an immune form, and
(3) a spatial extend form. Based on the analysis on its unique stochastic
differential equation and relevant Fokker-Planck equation, we will show that
the spatiotemporal fluctuation can lead to a transition of tumor-growth
state through both theoretical analysis and numerical calculation. Although
noise-induced phase transition is a well known phenomenon, double-faces
effect of a noise on a tumor system have not been reported. Here we will
show how this transition affects the tumor-growth and how the effect depends
on the initial state of tumor. Our results are inconsistent with the
zero-dimensional reports that suggest the fluctuation of growth rate always
puts the tumor at a disadvantage [10, 11].

The tumor-growth under immune surveillance can be described by means of
insect outbreak model [1, 16, 17], which in non-dimensional units is given by%
\begin{equation}
\frac{du}{dt}=ru(1-\frac{u}{K})-\frac{\beta u^{2}}{1+u^{2}}
\end{equation}%
where $u$ is the population of tumor cells; $r$ is their linear per capita
birth rate and $K$ is the carrying capacity of the environment,
respectively. $\beta u^{2}/(1+u^{2})$ quantifies the abilities of immune
cells to recognize and attack tumor cells. In general, chemotherapy can lead
to a fluctuation of tumor growth, simply a fluctuation of tumor growth rate $%
r$. If considering the space of tumor-growth, the growth rate $r$ in Eq.(1)
should be rewritten as $r_{0}+\xi _{i}(t),$ where $\xi _{i}(t)$ is the
Gaussian noises, white in time and space, with zero mean and autocorrelation
defined by $\langle \xi _{i}(t)\rangle =0,\ \ \langle \xi _{i}(t)\xi
_{j}(t^{\prime })\rangle =2\sigma ^{2}\delta _{i,j}\delta (t-t^{\prime })$,
in which $\sigma ^{2}$ is the noise level and $i,j$ are lattice sites. The
equivalent stochastic differential equation of Eq.(1) will be,%
\begin{eqnarray}
\frac{du_{i}}{dt} &=&r_{0}u_{i}(1-\frac{u_{i}}{K})-\frac{\beta u_{i}^{2}}{%
1+u_{i}^{2}}+u_{i}(1-\frac{u_{i}}{K})\xi _{i}(t)  \notag \\
&&-\frac{D}{2d}\sum_{j\epsilon n(i)}(u_{i}-u_{j})
\end{eqnarray}%
here $n(i)$ is the set of the $2d$ nearest neighbors of site $i$, $d$ and $D$
are the spatial dimension and the diffusion coefficient, respectively.

Equations of this kind are general and cover different tumor growth and
diffusion phenomena, especially nonequilibrium growth. We would like to
track down the existence of nonequilibrium phase transition induced by
multiplicative noise, in systems described by these equations. Such a phase
transition is characterized by the appearance of multiple steady state
probability distributions $p_{st}(\{u_{i}\}),$ which has been applied
successfully in numerous stochastic problems [18, 19]. If set $%
f(u_{i})=r_{0}u_{i}(1-u_{i}/K)-\beta u_{i}^{2}/(1+u_{i}^{2}),$ and $%
g(u_{i})=u_{i}(1-u_{i}/K),$ one will obtain the equivalent Fokker-Planck
equation of Eq.(2),%
\begin{equation}
\frac{\partial p(\{u_{i}\},t)}{\partial t}=-\frac{\partial \lbrack
A(u_{i})p(\{u_{i}\},t)]}{\partial u_{i}}+\frac{\partial
^{2}[B(u_{i})p(\{u_{i}\},t)]}{\partial u_{i}^{2}}
\end{equation}%
in which%
\begin{eqnarray}
A(u_{i}) &=&f(u_{i})+\sigma ^{2}g(u_{i})g^{^{\prime }}(u_{i})+\frac{D}{2d}%
\sum_{j\epsilon n(i)}(u_{i}-u_{j})  \notag \\
B(u_{i}) &=&\sigma ^{2}g^{2}(u_{i})
\end{eqnarray}

For simplicity of notation, we drop the subscript $i$. The stationary
solution to Eq.(3) is given to be%
\begin{equation}
p_{st}(u)=Z\exp [\frac{2}{\sigma ^{2}}\int_{0}^{u}dv\frac{f(u)-\frac{\sigma
^{2}}{2}g(u)g^{^{\prime }}(u)-D[v-E(v)]}{g^{2}(u)}]
\end{equation}%
where $Z$ is a normalization constant, and 
\begin{equation}
E(v)=\langle v_{i}|v_{j}\rangle =\int v_{j}p_{st}(v_{j}|v_{i})dv_{j},
\end{equation}%
represents the steady state conditional average of $v_{j}$ at neighboring
sites $j\in n(i)$, given the value $v_{i\text{ }}$at site $i$.

Using the Weiss mean field approximation [20, 21], neglecting the
fluctuation in the neighboring sites, i.e., $E(v)=\langle u\rangle $,
independent of $v$, and imposing the self-consistent requirement $m=\langle
u\rangle $, we obtain 
\begin{equation}
m=\frac{\int_{0}^{+\infty }up_{st}(u,m)du}{\int_{0}^{+\infty }p_{st}(u,m)du}%
=F(m)
\end{equation}

\ifcase\msipdfoutput
\FRAME{itpFU}{4.1857in}{3.3719in}{0in}{\Qcb{The solution, $m$, of the
self-consistency equation is the intersection point between $F(m)=m$ and $%
F(m)=y(m)$ for noise level $\sigma ^{2}=8.0\times 10^{3}.$}}{}{fig1.eps}{%
\special{language "Scientific Word";type "GRAPHIC";maintain-aspect-ratio
TRUE;display "ICON";valid_file "F";width 4.1857in;height 3.3719in;depth
0in;original-width 4.1355in;original-height 3.3252in;cropleft "0";croptop
"1";cropright "1";cropbottom "0";filename '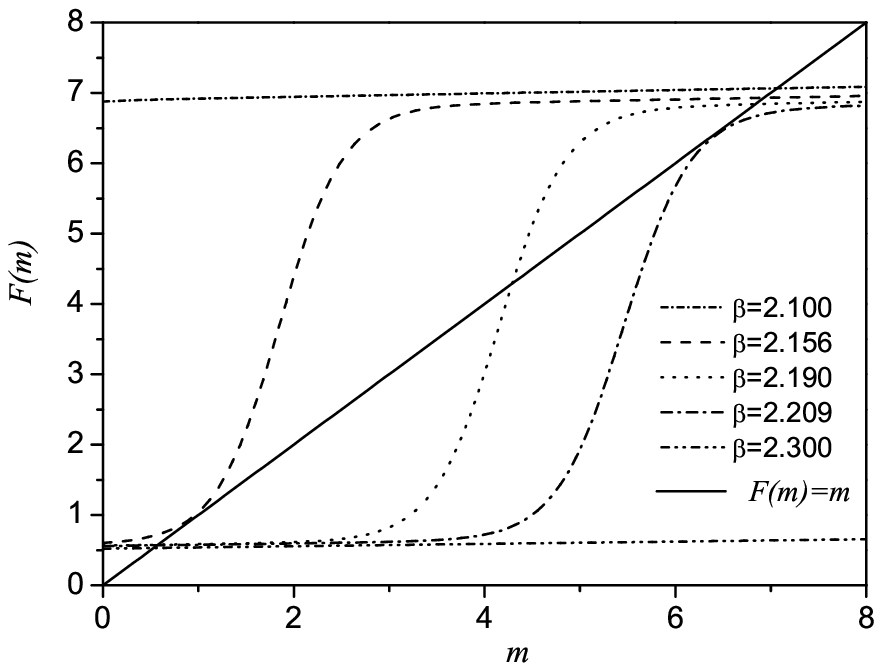';file-properties
"XNPEU";}}%
\else
\parbox[b]{4.1857in}{\begin{center}
\includegraphics[
natheight=3.3252in, natwidth=4.1355in, height=3.3719in, width=4.1857in]
{E:/temp/papers/graphics/fig1__1.pdf}\\
The solution, $m$, of the self-consistency equation is the intersection
point between $F(m)=m$ and $F(m)=y(m)$ for noise level $\sigma
^{2}=8.0\times 10^{3}.$%
\end{center}}%
\fi

\ifcase\msipdfoutput
\FRAME{ftpFU}{4.2004in}{3.442in}{0pt}{\Qcb{$m$ as a function of $\sigma ^{2}$
given by Eq.(7). The points correspond to the intersection of curves in
Fig.1. The critical immune coefficients are $\beta _{c1}=2.156$ and $\beta
_{c2}=2.209,$ respectively, which divide the state of a tumor into three
levels: excited (E), sub-excited (S) and non-excited (N).}}{}{fig2.eps}{%
\special{language "Scientific Word";type "GRAPHIC";maintain-aspect-ratio
TRUE;display "ICON";valid_file "F";width 4.2004in;height 3.442in;depth
0pt;original-width 4.1502in;original-height 3.3961in;cropleft "0";croptop
"1";cropright "1";cropbottom "0";filename '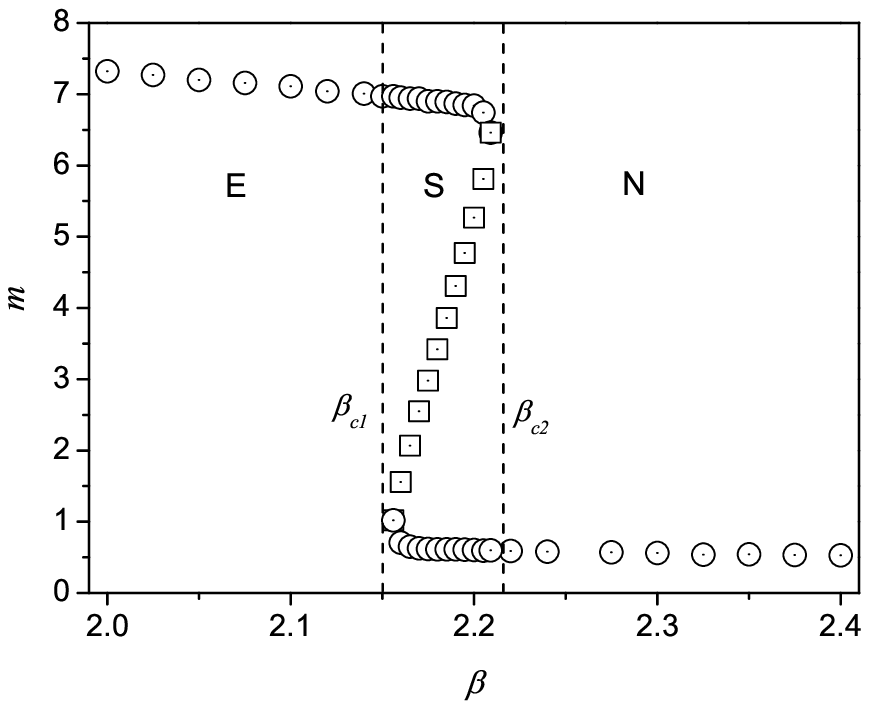';file-properties
"XNPEU";}}%
\else
\begin{figure}[pt]\begin{center}
\includegraphics[
natheight=3.3961in, natwidth=4.1502in, height=3.442in, width=4.2004in]
{E:/temp/papers/graphics/fig2__2.pdf}\caption{$m$ as a function of $\protect%
\sigma ^{2}$ given by Eq.(7). The points correspond to the intersection of
curves in Fig.1. The critical immune coefficients are $\protect\beta %
_{c1}=2.156$ and $\protect\beta _{c2}=2.209,$ respectively, which divide the
state of a tumor into three levels: excited (E), sub-excited (S) and
non-excited (N).}
\end{center}\end{figure}%
\fi

The numerical solution of this last equation for parameter values $%
r_{0}=1.0, $ $D=0.01,$ and $\sigma ^{2}=8.0\times 10^{-3}$ is shown in
Figs.1 and 2. The solution, $m$, as a function of immune coefficient, $\beta
,$ is obtained by the intersection point between $F(m)=m$ and $F(m)=y(m)$
(here $y(m)$ represents the function in the middle position of Eq.(7)).
Obviously, the average populations of tumor cells exhibit monostable state
for low and high values of $\beta $, but un-stable state for intermediate
value of $\beta $. The critical points are $\beta _{c1}=2.156$ and $\beta
_{c2}=2.209,$ which divide the states of tumor into three levels: excited
state (E), sub-excited state (S), and non-excited state (N). Here E and N
correspond to stable states but S represents an un-stable state, which has
two or three possible values. This result means the state of tumor is
determined by the immune coefficient for low value of noise intensity.

\ifcase\msipdfoutput
\FRAME{ftpFU}{4.7876in}{3.525in}{0pt}{\Qcb{Stationary probability
distributions of average population of tumor cells for different noise
intensities and immune coefficients. The parameters are (a) $\beta
=2.12,\sigma ^{2}=0.01$, (b) $\beta =2.30,\sigma ^{2}=0.01$, (c) $\beta
=2.12,\sigma ^{2}=0.40$, (d) $\beta =2.30,\sigma ^{2}=0.40$.}}{}{fig3.eps}{%
\special{language "Scientific Word";type "GRAPHIC";maintain-aspect-ratio
TRUE;display "ICON";valid_file "F";width 4.7876in;height 3.525in;depth
0pt;original-width 4.734in;original-height 3.4783in;cropleft "0";croptop
"1";cropright "1";cropbottom "0";filename '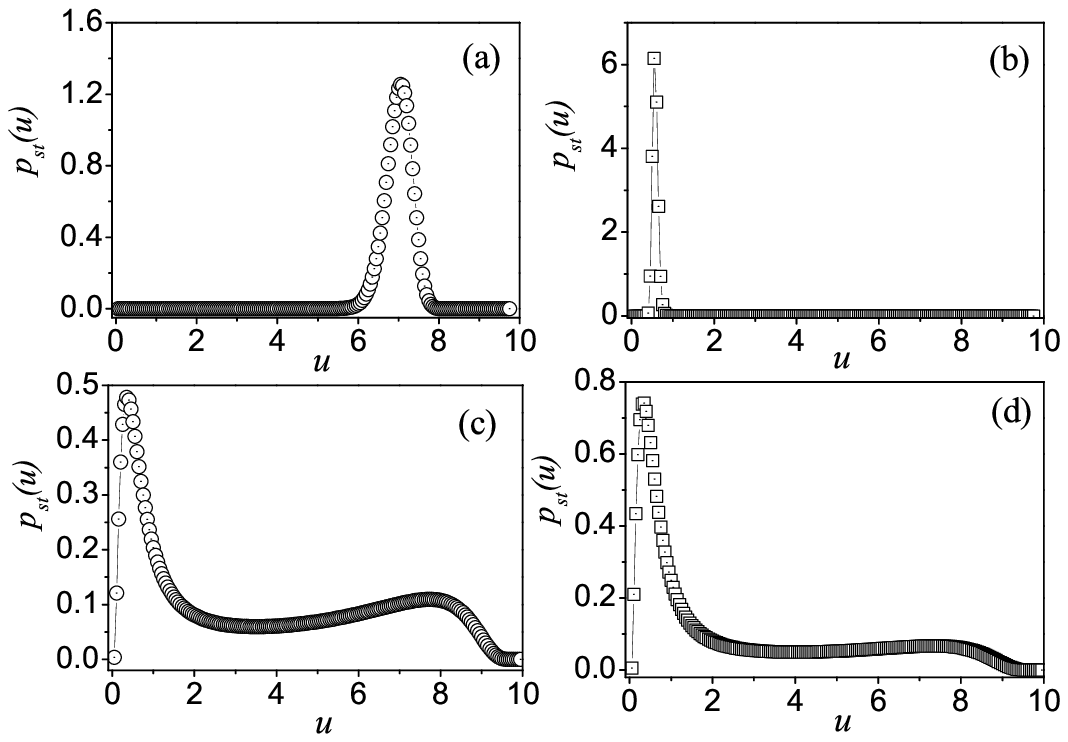';file-properties
"XNPEU";}}%
\else
\begin{figure}[pt]\begin{center}
\includegraphics[
natheight=3.4783in, natwidth=4.734in, height=3.525in, width=4.7876in]
{E:/temp/papers/graphics/fig3__3.pdf}\caption{Stationary probability
distributions of average population of tumor cells for different noise
intensities and immune coefficients. The parameters are (a) $\protect\beta %
=2.12,\protect\sigma ^{2}=0.01$, (b) $\protect\beta =2.30,\protect\sigma %
^{2}=0.01$, (c) $\protect\beta =2.12,\protect\sigma ^{2}=0.40$, (d) $\protect%
\beta =2.30,\protect\sigma ^{2}=0.40$.}
\end{center}\end{figure}%
\fi

\ifcase\msipdfoutput
\FRAME{ftpFU}{4.1857in}{3.3996in}{0pt}{\Qcb{$m$ as a function of $\sigma ^{2}
$ given by Eq.(7). The points are obtained by a method as same as for Fig.2.}%
}{}{fig4.eps}{%
\special{language "Scientific Word";type "GRAPHIC";maintain-aspect-ratio
TRUE;display "ICON";valid_file "F";width 4.1857in;height 3.3996in;depth
0pt;original-width 4.1355in;original-height 3.3529in;cropleft "0";croptop
"1";cropright "1";cropbottom "0";filename '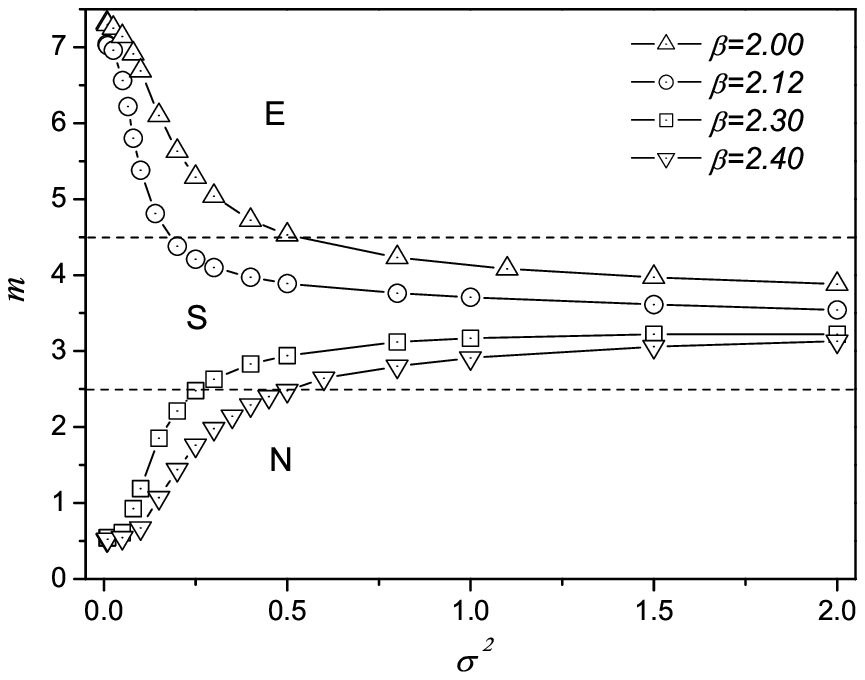';file-properties
"XNPEU";}}%
\else
\begin{figure}[pt]\begin{center}
\includegraphics[
natheight=3.3529in, natwidth=4.1355in, height=3.3996in, width=4.1857in]
{E:/temp/papers/graphics/fig4__4.pdf}\caption{$m$ as a function of $\protect%
\sigma ^{2}$ given by Eq.(7). The points are obtained by a method as same as
for Fig.2.}
\end{center}\end{figure}%
\fi

When the noise level $\sigma ^{2}$ increases, what will happen? To answer
this question, we consider E and N, respectively, shown in Fig.3, the
stationary probability distributions $p_{st}(u)$ change from monostable
state to bistable state with increasing noise intensity and more
quantitative results are given by Fig.4. For a tumor with excited state,
shown in Fig.4, when noise level increase, its growth can be hold back to a
sub-excited state. Conversely, for the non-excited tumor, noise can lead the
tumor to sub-excited state or even cancerization. This theoretical results
are confirmed by corresponding simulations of a one-dimensional system,
shown in Fig.5, obtained through a numerical integration of the set of
stochastic differential equations (2) [22, 23]. In the simulation, we
consider three sizes but not find one-dimensional finite size effect. It is
an important future work to analyze multi-dimensional phase transition of
tumor system in such a homogeneous circumstance.

\ifcase\msipdfoutput
\FRAME{ftpFU}{4.2004in}{3.442in}{0pt}{\Qcb{One dimensional simulation for
the relationship between $m$ and $\sigma ^{2}$. The parameters are same as
for Fig.4 }}{}{fig5.eps}{%
\special{language "Scientific Word";type "GRAPHIC";maintain-aspect-ratio
TRUE;display "ICON";valid_file "F";width 4.2004in;height 3.442in;depth
0pt;original-width 4.1502in;original-height 3.3961in;cropleft "0";croptop
"1";cropright "1";cropbottom "0";filename '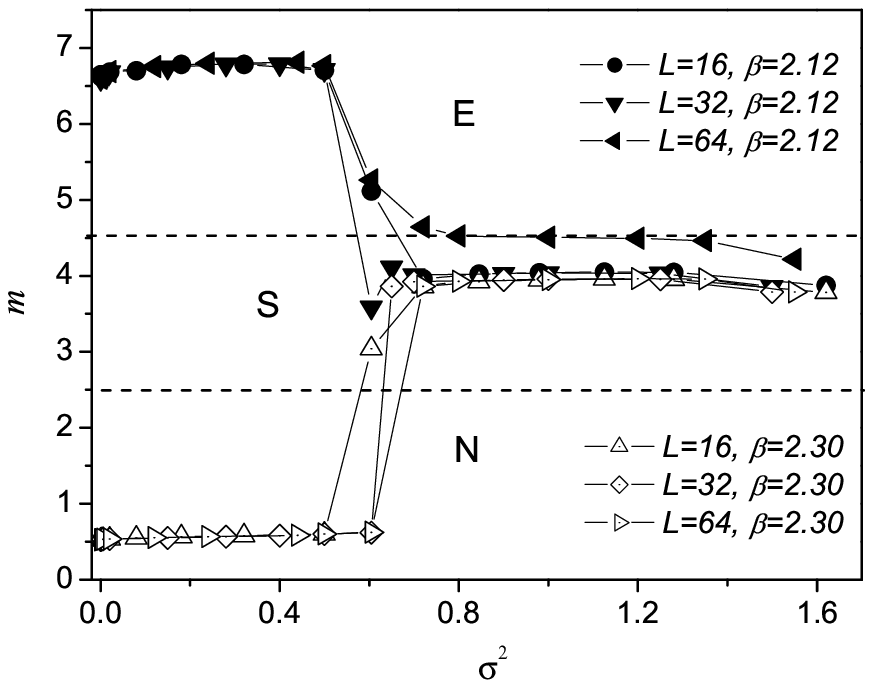';file-properties
"XNPEU";}}%
\else
\begin{figure}[pt]\begin{center}
\includegraphics[
natheight=3.3961in, natwidth=4.1502in, height=3.442in, width=4.2004in]
{E:/temp/papers/graphics/fig5__5.pdf}\caption{One dimensional simulation for
the relationship between $m$ and $\protect\sigma ^{2}$. The parameters are
same as for Fig.4 }
\end{center}\end{figure}%
\fi

In conclusion, we have found strong evidence for the existence of a
noise-induced different nonequilibrium phase transitions of tumor growth, in
which whether the noise advantage the tumor depends on the initial state of
tumor. When the tumor is excited, noise induces a decay. On the contrary, if
the tumor is inactive, the noise can stimulate its growth. Provided that the
noise results from the treatment as chemotherapy, our results suggest that
estimating the state of a tumor is a crucial work just before treatment
begins.

This work was partially supported by the National Natural Science Foundation
of China (Grant No. 60471023).

\end{document}